\pdfoutput=1
\documentclass[11pt,twoside,a4paper,cmspaper,final,collab]{cms-tdr}
\def\svnVersion{dee449e}\def\svnDate{2024/01/25}\def\cmsCernNoTag{CERN-EP-2023-083}\def\cmsCernDate{\today}\def\cmsMessage{Published in Physical Review Letters as \href{http://dx.doi.org/10.1103/PhysRevLett.132.041802}{\doi{10.1103/PhysRevLett.132.041802}.}}
\begin{document}\cmsNoteHeader{EXO-20-010}

\DeclareRobustCommand{\Ap}{{\HepParticle{A}{}{\prime}}\Xspace}
\newcommand{\dm}{\ensuremath{\chi_1}\xspace}
\newcommand{\DM}{\ensuremath{\chi_2}\xspace}
\newcommand{\mdm}{\ensuremath{m_1}\xspace}
\newcommand{\mDM}{\ensuremath{m_2}\xspace}
\newcommand{\msplit}{\ensuremath{\Delta}\xspace}
\newcommand{\mAp}{\ensuremath{m_{\Ap}}\xspace}
\newcommand{\ctau}{\ensuremath{c\tau}\xspace}
\newcommand{\dphi}{\ensuremath{\Delta\phi_{\PGm\PGm}^{\text{miss}}}\xspace}
\newcommand{\aD}{\ensuremath{\alpha_{\text{D}}}\xspace}
\newcommand{\aEM}{\ensuremath{\alpha_{\text{EM}}}\xspace}
\newcommand{\kinmix}{\ensuremath{\epsilon}\xspace}
\newcommand{\dxy}{\ensuremath{d_{xy}}\xspace}
\newcommand{\vxy}{\ensuremath{v_{xy}}\xspace}
\newcommand{\mindxy}{\ensuremath{\text{min-}\dxy}\xspace}
\newcommand{\pfiso}{\ensuremath{I_{\text{PF}}^{\text{rel}}}\xspace}
\newcommand{\cmsTableAlways}[1]{\resizebox{\textwidth}{!}{#1}}
\newlength\cmsTabSkip\setlength{\cmsTabSkip}{0.3ex}
\ifthenelse{\boolean{cms@external}}{\newcommand{\cmsTablePRL}[1]{\resizebox{\columnwidth}{!}{#1}}}{\newcommand{\cmsTablePRL}[1]{#1}}
\ifthenelse{\boolean{cms@external}}{\newcommand{\cmsLeft}{upper\xspace}}{\newcommand{\cmsLeft}{left\xspace}}
\ifthenelse{\boolean{cms@external}}{\newcommand{\cmsRight}{lower\xspace}}{\newcommand{\cmsRight}{right\xspace}}
\ifthenelse{\boolean{cms@external}}{\providecommand{\prlWidth}{0.95\columnwidth}}{\providecommand{\prlWidth}{0.56\textwidth}}

\cmsNoteHeader{EXO-20-010}

\title{Search for inelastic dark matter in events with two displaced muons and missing 
transverse momentum in proton-proton collisions at \texorpdfstring{$\sqrt{s} = 13\TeV$}{sqrt(s) = 13 TeV}}

\date{\today}

\abstract{
A search for dark matter in events with a displaced nonresonant muon pair and missing transverse momentum is presented. The analysis is performed using an integrated luminosity of 138\fbinv of proton-proton ($\Pp\Pp$) collision data at a center-of-mass energy of 13\TeV produced by the LHC in 2016--2018. No significant excess over the predicted backgrounds is observed. Upper limits are set on the product of the inelastic dark matter production cross section $\sigma(\Pp\Pp\to \Ap \to \dm \, \DM)$ and the decay branching fraction $\mathcal{B}(\DM \to \dm \, \PGmp \, \PGmm)$, where \Ap is a dark photon and \dm and \DM are states in the dark sector with near mass degeneracy. This is the first dedicated collider search for inelastic dark matter.
}

\hypersetup{%
pdfauthor={CMS Collaboration},%
pdftitle={Search for inelastic dark matter in events with two displaced muons and missing transverse momentum in proton-proton collisions at sqrt(s) = 13 TeV},%
pdfsubject={CMS},%
pdfkeywords={CMS, exotica, long-lived, inelastic, dark matter, dark photon, lepton-jet}}

\maketitle

The presence of dark matter (DM) is strongly supported by observations~\cite{ARBEY2021103865,Clowe_2004,2009ApJS..180..330K,PhysRevLett.115.231301,Planck:2018vyg}, but its nature remains largely unknown. Dedicated experiments (\eg, Refs.~\cite{Schumann_2019,PhysRevLett.118.021303,PhysRevD.103.063028,SuperCDMS:2017mbc}) have searched for DM directly, but no signal has yet been detected. 
Particle colliders are a complementary tool in this effort. Several searches for minimal models of DM have been carried out at the CERN LHC, such as those predicting weakly interacting massive particles~\cite{ATLASConstraintsMediatorbasedDark2019,ATLAS_Monojet2021,CMS_Monojet2021,ATLAS_MonoHiggs2021,Atlas_HiggsInv2023,CMS_HiggsInv2023}. Collider-based searches for long-lived particles (LLPs) can probe a wider range of DM models than previously explored~\cite{ATLAS:2018rjc,ATLAS:2019fwx,Alimena_2020,ATLAS_LLP_leptons,sirunyanSearchDecaysStopped2018,LHCb:2020ysn,PhysRevLett.124.131802,CMS_displacedJets,scoutingdispdimuon2022,EXO-18-003,EXO-21-006}. These particles can travel macroscopic distances before decaying inside the detector, leaving unique signatures. Several theoretical mechanisms predict a suppressed phase space for the production and decay of DM states, which would lead to long-lived DM phenomenology at the LHC~\cite{Alimena_2020}. Moreover, targeting LLPs has the considerable advantage of reducing or even eliminating a large class of standard model (SM) backgrounds, thus improving sensitivity for models with low-energy final-state particles, a theoretically well-motivated but typically challenging signature~\cite{PhysRevD.43.3191,Edsj__1997,izaguirre,berlinInelasticDarkMatter2019}.

In this Letter, we present a novel search for DM production in LHC proton-proton ($\Pp\Pp$) collisions that targets displaced-decay signatures. Specifically, we probe an inelastic DM (IDM) model~\cite{Tucker-Smith:2001myb,izaguirre,berlinInelasticDarkMatter2019} that postulates the existence of at least two states in the dark sector (the lightest of which is the stable DM) accompanied by a dark photon that kinetically mixes with the SM hypercharge~\cite{Holdom:1985ag}. In inelastic coupling scenarios, these states cannot scatter elastically with other particles (\eg, nucleons)~\cite{Tucker-Smith:2001myb}. These models predict a flavor-mixing off-diagonal vector coupling between the states and the dark photon, such that both states are produced simultaneously. A small mass gap between the lighter and heavier states leads to a compressed phase space and increased lifetime of the heavier state, thus producing LLP decay signatures in the CMS detector. Such models can account for the observed thermal-relic abundance~\cite{izaguirre}, which is the density of dark matter energy left over from the early evolution of the universe, while evading the increasingly stringent experimental constraints set by other DM searches. Previous studies~\cite{izaguirre,Batell_2014,PhysRevD.96.055007,Mongillo_2023} have placed bounds on the IDM production cross section for lighter dark matter masses (${<} 1\GeV$) by reinterpreting existing results from previous experiments~\cite{PhysRevD.38.3375,Auerbach_2001,Hook:2010tw,Lees_2017} or via fixed-target experimental setups. The work described here is the first dedicated collider search for IDM, which provides new sensitivity to heavier DM masses (${\approx}3$--80\GeV) and to displaced nonresonant dimuon production. This is achieved via the use of a dedicated displaced muon reconstruction algorithm and optimized event selection criteria, including isolation requirements. The signal selection efficiencies may be as low as $10^{-4}$ for low DM masses and highly displaced signal hypotheses, but the predicted IDM production cross sections can be as large as a few picobarns. The pioneering sensitivity achieved in this work to the unique final-state topology comprising low-\pt, displaced, and nonresonant muon pairs foreshadows the increasing attention devoted to more complex scenarios of new physics, given the absence of conclusive evidence for resonances or other singular phenomena incompatible with the SM. The techniques presented here will enable novel searches targeting models of DM with a rich structure and other compressed-spectra models such as supersymmetry models featuring coannihilation~\cite{Baer_2011,PhysRevD.89.075007}. Tabulated results are provided in the HEPData record for this analysis~\cite{hepdata}. 

\begin{figure}[hbt]
    \centering
    \includegraphics[width=0.99\columnwidth]{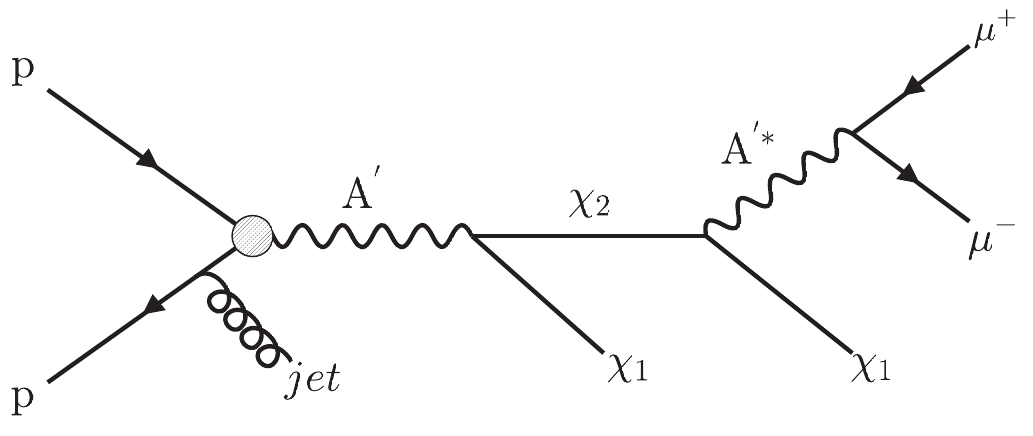}
    \caption{Feynman diagram of IDM production and decay in $\Pp\Pp$ collisions. The heavier state \DM can be long-lived, and decays to \dm and to a muon pair via an off-shell dark photon \Ap.}
    \label{fig:iDM_signature}
\end{figure}

The process of interest includes a dark photon \Ap that is produced in the $\Pp\Pp$ collision and recoils from an initial-state radiation jet, which is needed for efficient triggering on the event. The \Ap decays promptly to two states \dm and \DM with a near mass degeneracy, as illustrated in Fig.~\ref{fig:iDM_signature}.
The light DM state \dm (with mass \mdm) is stable and not detected. The heavy state \DM (with mass \mDM) travels a measurable distance before decaying to another \dm and a pair of SM particles via an off-shell dark photon. The mass splitting between the two states $\msplit \equiv \mDM - \mdm$ is relatively small, between 10--40\% of \mdm, resulting in decay products that are ``soft" (transverse momentum $\pt \lesssim 15\GeV$) and have small angular separation. Here we focus on final states with muons because of the powerful reconstruction and identification tools available for displaced muons and the higher purity of the reconstruction compared to that of soft electrons. The escaping \dm particles are collimated with the soft muons and lead to sizable \ptvecmiss, defined as the projection onto the plane perpendicular to the beam axis of the negative vector momentum sum of all reconstructed objects in an event. Its magnitude is referred to as \ptmiss. Previous CMS, ATLAS, and LHCb searches for processes with dimuon signatures~\cite{ATLAS:2018rjc,ATLAS:2019fwx,LHCb:2020ysn,PhysRevLett.124.131802,scoutingdispdimuon2022,EXO-21-006,sirunyanSearchDecaysStopped2018,atlascollaboration2022search,scoutingdispdimuon2022,LHCb:2020ysn} are not as sensitive to this class of models, for various reasons. Some analyses rely on the detection of dimuon resonances~\cite{ATLAS:2018rjc,PhysRevLett.124.131802,scoutingdispdimuon2022,EXO-21-006}, whereas muons in IDM are not resonantly produced. Others must compromise on either requiring higher muon \pt~\cite{sirunyanSearchDecaysStopped2018,ATLAS:2019fwx,atlascollaboration2022search,EXO-21-006} to keep the background and trigger rates manageable, or requiring muons to be prompt~\cite{scoutingdispdimuon2022,LHCb:2020ysn}. However, such selections also remove a significant fraction of IDM signal-like events, which feature both soft and displaced muons. Finally, searches for monojet signatures could have some sensitivity to IDM models for muon displacements sufficiently high such that both muons are not reconstructed, but there would be a large background from multijet processes, as well as low signal production at the high muon-pair displacement. The signal sensitivity would therefore be small. The required signature including displaced muons allows backgrounds to be controlled without the need for a higher \ptmiss threshold, which would greatly reduce the efficiency to reconstruct the IDM signal.
 
The central feature of the CMS apparatus is a superconducting solenoid of 6\unit{m} internal diameter, providing 
a magnetic field of 3.8\unit{T}. The solenoid volume contains a silicon pixel and strip tracker (extending 
in radius from 4 to 110\unit{cm} and covering the range of pseudorapidity $\abs{\eta} < 2.4$), a lead tungstate 
crystal electromagnetic calorimeter, and a brass-scintillator hadron calorimeter, each composed of a barrel 
and two endcap sections. Forward calorimeters extend the $\eta$ coverage provided by the barrel and endcap 
detectors. Muons are detected in gas-ionization detectors embedded in the steel flux-return yoke outside 
the solenoid, extending radially from about 4 to 7\unit{m} and covering the range $\abs{\eta} < 2.4$. A more 
detailed description of the CMS detector, together with a definition of the coordinate system used and 
the relevant kinematic variables, can be found in Ref.~\cite{Chatrchyan:2008zzk}. A particle-flow (PF) algorithm~\cite{CMS-PRF-14-001} 
aims to reconstruct and identify each individual particle in an event, with an optimized combination of 
information from the various elements of the CMS detector.

The analysis is carried out with data collected by the CMS Collaboration in 2016--2018 with a total integrated 
luminosity of 138\fbinv. Simulated samples of signal and background events are used to optimize the event selection.
Signal samples with dimuon decays are generated with \MGvATNLO v.2.6.0~\cite{MadGraph,Alwall_2007} at leading 
order in quantum chromodynamics (QCD) and injected into \PYTHIA v8.226~\cite{Sjostrand:2014zea} for fragmentation and parton shower modeling. 
Motivated by Ref.~\cite{izaguirre} and other sources, we select parameters consisting of \mdm in the range 3--80\GeV,
$\mAp = 3 \, \mdm$, $\Delta = \{0.1, 0.4\} \mdm$, \ctau in the range 1--1000\mm, 
and $\aD = \{0.1, 0.4\}$. Here, \ctau is the proper decay length of \DM and \aD 
is the coupling strength of the $U(1)_\mathrm{D}$ in the dark sector. These five parameters fix the kinetic 
mixing coefficient \kinmix, which controls the amount of mixing between the dark photon and the SM hypercharge. The CUETP8M1 
underlying event tune~\cite{khachatryanEventGeneratorTunes2016} is applied to 2016 samples and the CP5 tune~\cite{sirunyanExtractionValidationNew2020} 
to 2017--2018 samples. Two parton distribution function sets are used: NNPDF3.0~\cite{ballPartonDistributionsLHC2015} (2016) 
and NNPDF3.1~\cite{ballPartonDistributionsHighprecision2017} (2017--2018). Additional $\Pp\Pp$ interactions 
in the same or adjacent bunch crossings (pileup) are also simulated, with a frequency distribution matching 
that in data. Finally, the detector response is simulated with \GEANTfour~\cite{Agostinelli:2002hh} and identical 
reconstruction algorithms are applied to collision and simulated samples.

The event selection exploits the unique features expected from IDM: a pair of collinear, soft, and displaced muons collimated with \ptvecmiss. The muons are too soft to pass the trigger selection, so at least one energetic jet is required to boost the DM particles and enhance the \ptmiss spectrum. Candidate events are first selected with triggers with a minimum \ptmiss threshold of 120\GeV at the trigger level and 200\GeV in the offline selection, constructed excluding muons~\cite{Khachatryan:2016bia}. 
The leading jet in the event is required to have $\pt > 80\GeV$ and $\abs{\eta} < 2.4$. To accommodate additional initial-state emissions, only one other jet, with $\pt > 30\GeV$ and $\abs{\eta} < 5.0$, is allowed per event. The jet \pt requirement removes low-momentum jets produced in pileup interactions. Limiting the number of jets reduces the dominant background from events with jets produced through the strong interaction, referred to as QCD events, while retaining approximately 70\% of the signal yield. To suppress top quark backgrounds, events are vetoed if any jets are identified as originating from a bottom quark, based on the loose working point of the \textsc{DeepCSV} algorithm~\cite{Sirunyan:2298594,CMS-DP-2023-005}. 
The leading (subleading) jet must be azimuthally separated from \ptvecmiss by at least 1.50 (0.75)\unit{rad}. These selections further suppress 
backgrounds by ensuring that the DM system is well isolated in the event. 

\begin{figure}[hbt]
    \centering
    \includegraphics[width=\prlWidth]{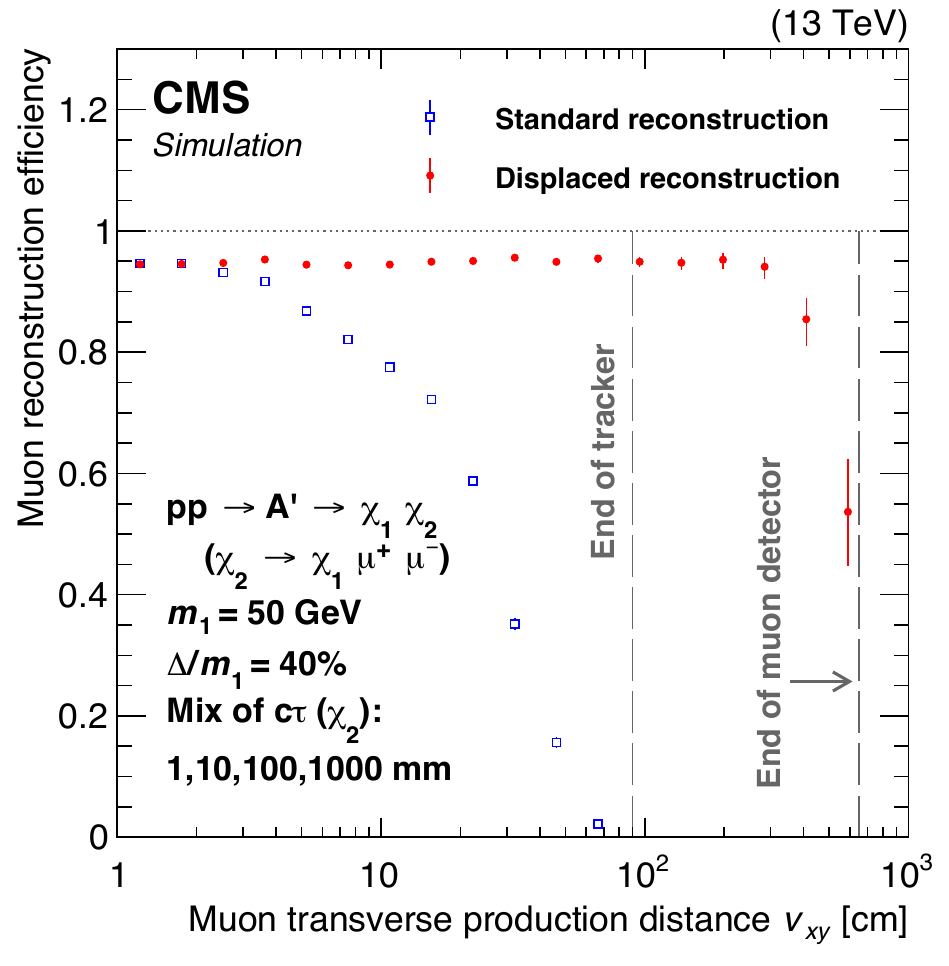}
    \caption{Simulated muon reconstruction efficiency of standard (blue squares) and displaced (red circles) reconstruction 
    algorithms as a function of transverse vertex displacement \vxy. The two dashed vertical gray lines denote 
    the ends of the fiducial tracker and muon detector regions, respectively.}
    \label{fig:reco_eff}
\end{figure}

Muons are reconstructed with a specialized algorithm designed to remain efficient even for large displacements 
of the muon-pair vertex of up to several meters from the luminous region. This displaced standalone algorithm (DSA) 
uses only information from the muon system and does not require muons to originate from the interaction 
point~\cite{Sirunyan_2018,sirunyanSearchDecaysStopped2018}. In Fig.~\ref{fig:reco_eff}, the DSA 
reconstruction efficiency for a representative signal sample is compared to that of the standard global 
reconstruction algorithm~\cite{Sirunyan_2018}, which requires both tracker and muon detector information. The 
efficiency is calculated \vs the distance \vxy in the transverse plane between the primary vertex (defined 
as the vertex with the largest value of summed charged particle $\pt^2$) and the muon-pair vertex. The 
efficiency remains high even when the muon-pair vertex lies beyond the inner radius of the muon detector planes.

The baseline muon selection requires at least two identified DSA muons. The identification criteria comprise ${>}12$ hits across ${\geq}2$ different muon detector planes (and ${>}18$ hits if no hits are found in the muon detector endcaps), track fit quality $\chi^2/\text{d.o.f} < 2.5$, \pt resolution $\sigma_{\pt}/\pt < 1$, $\pt > 5\GeV$, and $\abs{\eta} < 2.4$. The efficiency to identify such a DSA muon (about 90\%) is measured with three different data samples providing complementary coverage of the kinematic phase space: cosmic ray muons (as a proxy for displaced muons), muon pairs from {\PZ} boson decays (high \pt), and muon pairs from {\JPsi} meson decays (low \pt). This efficiency is compared to that of the corresponding simulated sample and the yearly efficiency ratio is parametrized as a function of muon \pt, $\eta$, and transverse impact parameter \dxy (defined as the closest distance in the transverse plane between the track trajectory and the main vertex), and applied as a correction to simulated events~\cite{EXO-21-006}. A cosmic ray muon veto is implemented by discarding events containing at least one pair of back-to-back DSA muons.
At least one pair of oppositely charged DSA muons must form a well-reconstructed vertex (using a Kalman filtering algorithm~\cite{fruhwirthApplicationKalmanFiltering1987a}) as inferred from a vertex fit with quality $\chi^2/\text{d.o.f} < 4$ and a relative position resolution of about 2-3\% (1\%) at low (high) displacements. The muon pair with the smallest $\chi^2$ value is chosen. The overall efficiency to correctly assign muon charges is greater than 97\% because of the low-energy spectrum of IDM muons. Finally, we require the muons to be collimated with $\Delta R \equiv \sqrt{\smash[b]{(\Delta\phi)^2 + (\Delta\eta)^2}} < 0.9$, and the dimuon \pt to be azimuthally aligned with the \ptmiss in the event, $\dphi \equiv \abs{\Delta\phi(\ptvecmiss, \vec{p}^{\PGm\PGm}_{\mathrm{T}})} < 0.5$. These requirements remove more than 90\% of the electroweak SM backgrounds without incurring a significant loss of signal efficiency.

The DSA muons have higher reconstruction efficiencies but degraded momentum resolution and impact parameter resolution compared to muons reconstructed with the standard algorithm. This is due to the smaller lever arm and lack of tracker information. To recover some of the performance at lower displacements, a match-and-replace procedure is implemented whereby DSA muon system tracks that are found close to muons reconstructed with the PF algorithm (PF muons) are replaced with the latter. The matching requirement is $\Delta R < 0.2$ at the location of the innermost muon spectrometer track hit. The baseline requirement of two DSA muons in the event selection removes ambiguities in the PF-DSA matching. The vertex is then refitted and again required to pass the $\chi^2/\text{d.o.f}<4$ threshold. We divide the signal region (SR) into three categories (0--2) based on the number of PF-DSA matches found. Highly displaced signal events fall in the zero-match category, since typically no PF muons are reconstructed, while signal events with slightly displaced muons fall in the one- or two-match categories, like most SM backgrounds. This categorization thus further enhances the search sensitivity for displaced signal models in the zero-match category.

Backgrounds passing the event selection consist mainly of three types: QCD events with genuine or misidentified muons and \ptmiss arising from jet mismeasurements (dominant in all match categories); multijet {\PW} events where the \PW boson decays to a muon and a muon-neutrino and a misidentified muon is selected (contributing to the one-match category); and multijet {\PZ} events where the \PZ boson decays to two neutrinos and two misidentified muons are selected (contributing to the zero- and one-match categories). Misidentified muons may be associated with pileup interactions or the underlying event and therefore not correlate with the hard scatter in the event, or they may arise from instrumental deficiencies. No backgrounds from cosmic ray muons or other noncollisional contributions are expected because of the highly selective event requirements, the cosmic ray veto, and the use of \ptmiss triggers. The absence of such backgrounds was checked with a dedicated sample of events collected with noncollisional triggers, which require particles traversing the CMS detector while simultaneously vetoing any beam particles crossing the down- and upstream ends of the detector.

A modified matrix (``$ABCD$") method is employed to estimate the backgrounds, relying on two independent 
variables to discriminate between signal and background. The two-dimensional plane formed by these variables 
is divided into four bins ($A$--$D$). Because the variables are independent, all four bins can be described 
using only three parameters: a normalization rate plus a vertical and a horizontal transfer coefficient. The 
final degree of freedom is then used to fit the signal rate across all bins~\cite{cmscollaborationSearchLonglivedParticles2019}.

To maximize sensitivity, different variables are used in each match category. All three match categories use the minimum \dxy of the two muons, referred to as \mindxy. In the one- and two-match categories, the second discriminating variable is the relative PF isolation $I^{\text{rel}}_{\text{PF}}$ of the \mindxy muon, defined as the \pt sum of all photons, charged and neutral hadrons found within a $\Delta R < 0.4$ cone of a muon, divided by its \pt. 
Unlike in QCD events, where muons originate from parton fragmentation and hadronization processes, muons in IDM should be isolated. In the zero-match category, where no PF muons are identified, the second $ABCD$ variable is \dphi. The assumption of independence between all variables is extensively checked, as described in the next paragraph. The simultaneous fit accounts for the presence of signal in all $ABCD$ bins, but a higher concentration is expected for high \mindxy, low \pfiso, and small \dphi.

To validate the background estimation procedure and to optimize the $ABCD$ binning, a multijet validation region (VR) enriched with backgrounds and devoid of signal is defined by inverting the requirement on the number of jets in the event, demanding at least three jets. Simulation studies show that QCD multijet events are the dominant background source in all match categories, comprising about 80\% of the expected backgrounds both in the SR and in the VR. The shapes of all observables used in the $ABCD$ procedure were found to agree within uncertainties between the VR and the SR, and between data and MC simulation. Several sets of $ABCD$ bins are defined in the VR and the agreement between predicted and observed yields in bin $D$ (the bin with the smallest backgrounds) is confirmed for each set. We choose a grid of bin boundaries centered around the optimal bins in each match category, for a total of 31 tests. The ensemble of control tests shows no statistically significant deviation between prediction and observation. Since these closure tests indicate that the observables chosen for the background validation are not correlated (within the statistical precision of the method), no systematic uncertainties are applied to the background prediction in the analysis, which is entirely dominated by statistical uncertainties. The optimal $ABCD$ binning (reported in Table~\ref{tab:results} for each match category) is determined by requiring at least three events in each of at least three bins of the $ABCD$ plane, which is needed for a reasonable background estimation in the fourth bin. 

\begin{table*}[htb]
\topcaption[ABCD bins considered per signal point]{Definition of $ABCD$ bins and yields in data, per match category. The predicted yield in the bin with the smallest backgrounds (bin $D$) is extracted from the simultaneous four-bin fit by assuming zero signal, which corresponds to $(\text{Obs. }B \times \text{Obs. }C) / (\text{Obs. }A)$ in this limit. Uncertainties in the predicted yields are purely statistical and determined from the propagation of Poisson uncertainties in the various bins.}
\label{tab:results}
\def\arraystretch{1.1}
\cmsTableAlways{
\centering\begin{scotch}{c|ccc|ccc|ccc}
\multirow{2}{*}{Bin} &                      \multicolumn{3}{c|}{Zero match}                        &                        \multicolumn{3}{c|}{One match}                       &                           \multicolumn{3}{c}{Two match}                     \\
                     & \dphi [rad]                  & \mindxy [cm]              & Events        & \pfiso                      & \mindxy [cm]                & Events        & \pfiso                      & \mindxy [cm]                & Events        \\
\hline
Obs. $A$               & \phantom{0.2}0--0.25         & 3--15                     & 68            & ${>}0.25$                   & 0.02--0.75                  & 716           & ${>}0.25$                   & 0.02--0.15                  & 424           \\
Obs. $B$               & 0.25--0.50                   & 3--15                     & 9             & ${<}0.25$                   & 0.02--0.75                  & 33            & ${<}0.25$                   & 0.02--0.15                  & 22            \\
Obs. $C$               & \phantom{0.2}0--0.25         & ${>}15$                   & 9             & ${>}0.25$                   & ${>}0.75$                   & 12            & ${>}0.25$                   & ${>}0.15$                   & 10            \\ [\cmsTabSkip]
Obs. $D$               & \multirow{2}{*}{0.25--0.50}  & \multirow{2}{*}{${>}15$}  & 2             & \multirow{2}{*}{${<}0.25$}  & \multirow{2}{*}{${>}0.75$}  & 0             & \multirow{2}{*}{${<}0.25$}  & \multirow{2}{*}{${>}0.15$}  & 0             \\
Pred. $D$              &                              &                           & $1.2 \pm 0.6$ &                             &                             & $0.5 \pm 0.2$ &                             &                             & $0.5 \pm 0.2$ 
\end{scotch}}
\end{table*}

The dominant signal uncertainties in the analysis are either statistical, arising from a low selection 
efficiency for some signal hypotheses, or systematic due to imperfect knowledge of efficiencies, energy 
corrections, and the integrated luminosity~\cite{Sirunyan:2759951,CMS-PAS-LUM-17-004,CMS-PAS-LUM-18-002}. The 
total signal systematic uncertainties averaged over all years are approximately 20, 30, and 40\% for 
the zero-, one-, and two-match categories, respectively (with a yearly breakdown shown in Table~\ref{tab:summary_uncert}). These 
are applied uniformly to all signal hypotheses, unlike the statistical uncertainty, which depends on the signal efficiency of each hypothesis. 

\begin{table}[htb]
\topcaption{Systematic uncertainties in the analysis. The jet uncertainties are larger in 2017 because of noise issues with the ECAL endcap. The tracking inefficiency in 2016 is caused by the unexpected saturation of photodiode signals in the tracker. The first two rows give the uncertainty per PF muon. Thus, for the first row, the contributions are 0\%, 5\%, and 10\% in the 0-, 1-, and 2-match categories, respectively. The third row lists the DSA displaced reconstruction (reco.) systematic uncertainty.}
\def\arraystretch{1.0}
\label{tab:summary_uncert}
\cmsTablePRL{
\centering\begin{scotch}{lcccc}
Uncertainty             & 2016      & 2017      & 2018      & Correlation   \\
\hline
PF displaced ID         & 5       & 5       & 5       & Total         \\
PF prompt ID            & 3.2     & 2.8     & 3.0     & Total         \\
DSA displaced reco.     & 2       & 2       & 2       & Total         \\
DSA displaced ID        & 2       & 2       & 2       & Total         \\
DSA prompt ID           & 0.6     & 0.7     & 0.6     & Total         \\
\PQb quark jet ID       & 0.5     & 0.5     & 0.5     & Total         \\
Electron \& photon ID   & 0.5     & 0.5     & 0.5     & Total         \\
Trigger                 & 1.5     & 1.5     & 1.5     & Total         \\
Jet energy resolution   & 1.0     & 9.0     & 2.5     & None          \\
Jet energy scale        & 2.0     & 6.0     & 2.0     & Total         \\
Luminosity              & 1.2     & 2.3     & 2.5     & Partial       \\
Tracking inefficiency   & 10      & \NA       & \NA       & \NA
\end{scotch}}
\end{table}

\begin{figure}[bt]
    \centering
    \includegraphics[width=\prlWidth]{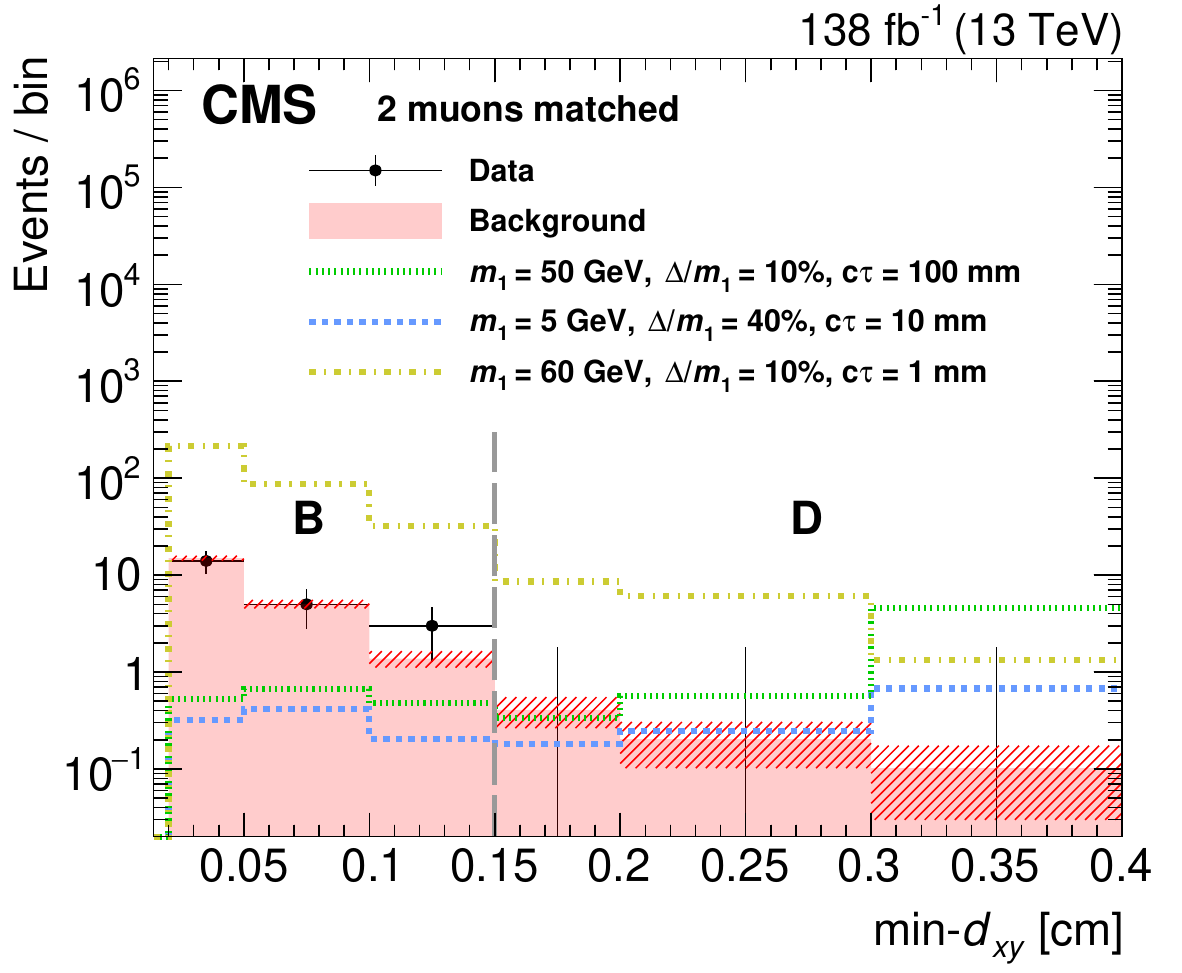}
    \caption{Measured \mindxy distribution in the two-match category, after requiring the \mindxy muon to pass the isolation 
    requirement $\pfiso <0.25$ (\ie, the $B$ and $D$ bins of the $ABCD$ plane). Overlaid with a red histogram is 
    the background predicted from the region of the $ABCD$ plane failing the same requirement (the $A$ and $C$ bins), 
    as well as three signal benchmark hypotheses (as defined in the legends), assuming \aD is equal to the electromagnetic (EM) fine-structure constant \aEM. The red hatched bands correspond to the background prediction uncertainty. The last bin includes the overflow.}
    \label{fig:min_dxy_2match}
\end{figure}

The observed yields in data are used to perform a simultaneous fit to the four $ABCD$ bins in each match category, and the fit results are found to be consistent with the background-only hypothesis, as illustrated in Table~\ref{tab:results}. Specifically, the measured (predicted) yields are 2 ($1.2 \pm 0.6$), 0 ($0.5 \pm 0.2$), and 0 ($0.5 \pm 0.2$) in the zero-, one-, and two-match categories, respectively. Figure~\ref{fig:min_dxy_2match} shows the measured \mindxy distribution in the two-match category, for the subset of events passing the isolation requirement ($B$ and $D$ bins). To derive the background prediction shown in Fig.~\ref{fig:min_dxy_2match}, the observed distribution in bins $A$ and $C$ is normalized to the sum of yields counted in bins $B$ and $D$ under the no-signal assumption.

\begin{figure}[bt]
    \centering
    \includegraphics[width=0.49\textwidth]{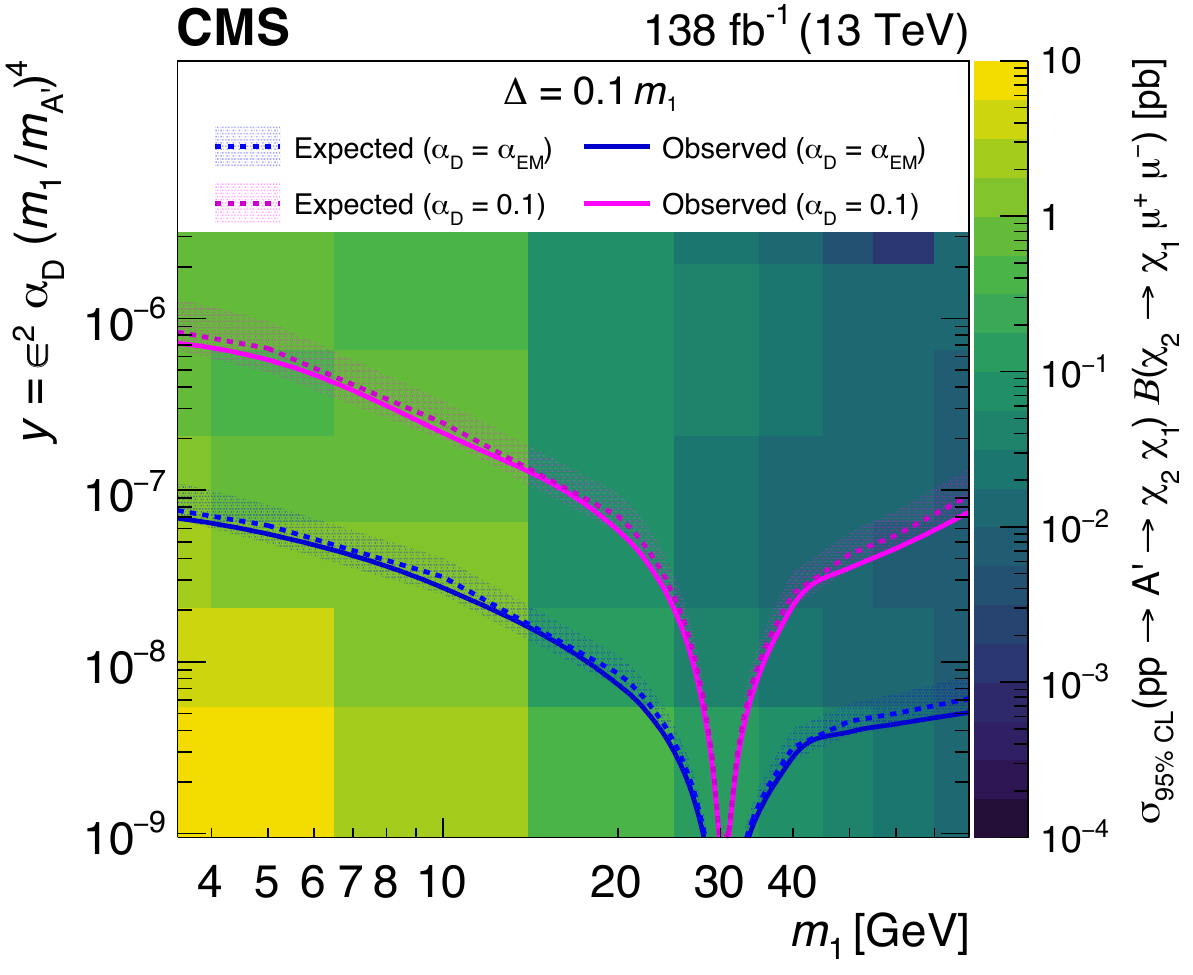}
    \includegraphics[width=0.49\textwidth]{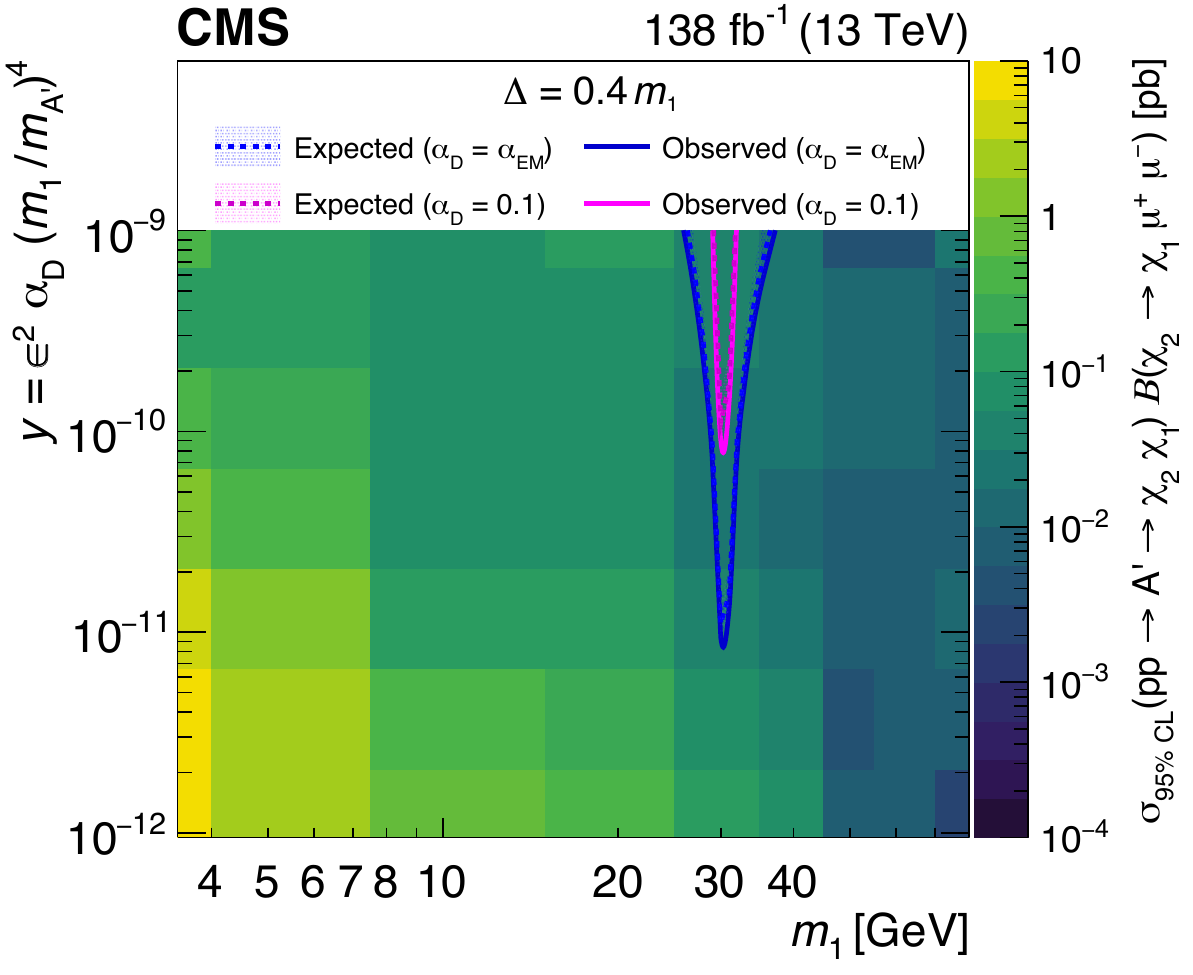}
    \caption{Two-dimensional exclusion surfaces for $\msplit = 0.1 \, \mdm$ (\cmsLeft) and $0.4 \, \mdm$ (\cmsRight), as functions of the DM mass \mdm and the signal strength $y$, with $\mAp = 3 \, \mdm$. Filled histograms denote observed limits on $\sigma(\Pp\Pp \to \Ap \to \DM \dm) \, \mathcal{B}(\DM \to \dm \PGmp \PGmm)$. Solid (dashed) curves denote the observed (expected) exclusion limits at 95\% CL, with 68\% CL uncertainty bands around the expectation. Regions above the curves are excluded, depending on the \aD hypothesis: $\aD = \aEM$ (dark blue) or 0.1 (light magenta). The sensitivity is higher in the region near $\mdm \approx 30 \GeV$ or $\mAp \approx 90\GeV$ because of the \Ap mixing with the \PZ boson in that mass range.}
    \label{fig:results}
\end{figure}

We assess 95\% confidence level (\CL) limits on the product of the DM production cross section and decay branching fraction $\sigma(\mathrm{pp}\to \Ap \to \DM \dm) \, \mathcal{B}(\DM \to \dm \PGmp \PGmm)$ using a modified frequentist criterion \CLs with the likelihood ratio test statistic~\cite{JUNK1999435,Read_2002}. The upper limits are shown in Fig.~\ref{fig:results} as a function of the mass \mdm of the DM state and the interaction strength $y \equiv \kinmix^2 \, \aD \, (\mdm/\mAp)^4$. This choice of parametrization allows the relevant variables to be scaled with the thermal-relic abundance in a straightforward way, as explored in Ref.~\cite{izaguirre}. The exclusion curves (in blue and magenta) depend more strongly on the choices of $\aD$ and of \mAp than the cross section limits because of the resonant enhancement arising from the mixing between the \Ap and \PZ bosons when $\mAp \approx m_\PZ$. Results are shown for $\mAp = 3 \, \mdm$, which is chosen for compatibility with the target thermal-relic DM abundances as determined by cosmological observations. Regions of parameter space are excluded in both mass-splitting scenarios: for $\msplit = 0.1 \, \mdm$, values of $y$ above ${\sim}10^{-7}$--$10^{-6}$ and above ${\sim}10^{-8}$--$10^{-7}$, depending on the choice of $\aD$, are excluded at $\mdm = 3$~and~80\GeV, respectively. For $\msplit = 0.4 \, \mdm$, the relative exclusion sensitivity is weaker because of the smaller production cross sections, which scale as $1/\msplit^5$ when all other parameters are fixed. Only the region near the \Ap-\PZ resonance is excluded in this scenario.

In summary, a search has been presented for inelastically coupled dark matter with a unique final-state 
signature including a soft, displaced muon pair collimated with the missing transverse momentum vector. The 
analysis is performed using proton-proton collision data produced by the LHC at a center-of-mass energy 
of 13\TeV and collected with the CMS experiment in 2016--2018. The data sample corresponds to an integrated 
luminosity of 138\fbinv. Control samples in data are used to predict the background, and no significant 
excess is observed over standard model expectations. Upper limits are set on the product of the DM production 
cross section and decay branching fraction into muons as a function of DM mass \mdm and interaction strength. This 
is the first dedicated collider search for inelastic dark matter and it significantly expands the sensitivity 
to \mdm above the GeV scale.

\begin{acknowledgments}
We thank Brian Shuve for providing calculations for some of the cross sections used in this Letter.

We congratulate our colleagues in the CERN accelerator departments for the excellent performance of the LHC and thank the technical and administrative staffs at CERN and at other CMS institutes for their contributions to the success of the CMS effort. In addition, we gratefully acknowledge the computing centers and personnel of the Worldwide LHC Computing Grid and other centers for delivering so effectively the computing infrastructure essential to our analyses. Finally, we acknowledge the enduring support for the construction and operation of the LHC, the CMS detector, and the supporting computing infrastructure provided by the following funding agencies: BMBWF and FWF (Austria); FNRS and FWO (Belgium); CNPq, CAPES, FAPERJ, FAPERGS, and FAPESP (Brazil); MES and BNSF (Bulgaria); CERN; CAS, MoST, and NSFC (China); MINCIENCIAS (Colombia); MSES and CSF (Croatia); RIF (Cyprus); SENESCYT (Ecuador); MoER, ERC PUT and ERDF (Estonia); Academy of Finland, MEC, and HIP (Finland); CEA and CNRS/IN2P3 (France); BMBF, DFG, and HGF (Germany); GSRI (Greece); NKFIH (Hungary); DAE and DST (India); IPM (Iran); SFI (Ireland); INFN (Italy); MSIP and NRF (Republic of Korea); MES (Latvia); LAS (Lithuania); MOE and UM (Malaysia); BUAP, CINVESTAV, CONACYT, LNS, SEP, and UASLP-FAI (Mexico); MOS (Montenegro); MBIE (New Zealand); PAEC (Pakistan); MES and NSC (Poland); FCT (Portugal); MESTD (Serbia); MCIN/AEI and PCTI (Spain); MOSTR (Sri Lanka); Swiss Funding Agencies (Switzerland); MST (Taipei); MHESI and NSTDA (Thailand); TUBITAK and TENMAK (Turkey); NASU (Ukraine); STFC (United Kingdom); DOE and NSF (USA).
\end{acknowledgments}

\bibliography{auto_generated}

\providecommand{\href}[2]{#2}\begingroup\raggedright\begin{thebibliography}{10}%
\makeatletter
\providecommand{\hrefCMSnoop }[0]{\@secondoftwo}%
\makeatother
\providecommand{\doi}{\texttt{doi:}\begingroup \urlstyle{tt}\Url}

\bibitem{ARBEY2021103865}
\hrefCMSnoop {}{A.~Arbey and F.~Mahmoudi, ``Dark matter and the early
  {Universe}: A review'',} \textit{ Prog. Part. Nucl. Phys.} \textbf{ 119}
  (2021) 103865,
  \href{http://dx.doi.org/10.1016/j.ppnp.2021.103865}{\doi{10.1016/j.ppnp.2021.103865}},
  \href{http://www.arXiv.org/abs/2104.11488}{\texttt{arXiv:2104.11488}}.

\bibitem{Clowe_2004}
\hrefCMSnoop {}{D.~Clowe, A.~Gonzalez, and M.~Markevitch, ``Weak-lensing mass
  reconstruction of the interacting cluster {1E} 0657-558: Direct evidence for
  the existence of dark matter'',} \textit{ Astrophys. J.} \textbf{ 604} (2004)
  596, \href{http://dx.doi.org/10.1086/381970}{\doi{10.1086/381970}},
  \href{http://www.arXiv.org/abs/astro-ph/0312273}{\texttt{arXiv:astro-ph/0312273}}.

\bibitem{2009ApJS..180..330K}
\hrefCMSnoop {}{E.~{Komatsu} {et~al.}, ``Five-year {Wilkinson Microwave
  Anisotropy Probe} observations: Cosmological interpretation'',} \textit{
  Astrophys. J. Suppl. Ser.} \textbf{ 180} (2009) 330,
  \href{http://dx.doi.org/10.1088/0067-0049/180/2/330}{\doi{10.1088/0067-0049/180/2/330}},
  \href{http://www.arXiv.org/abs/0803.0547}{\texttt{arXiv:0803.0547}}.

\bibitem{PhysRevLett.115.231301}
\hrefCMSnoop {}{{Fermi-LAT} Collaboration, ``Searching for dark matter
  annihilation from {M}ilky {W}ay dwarf spheroidal galaxies with six years of
  {Fermi} {Large} {Area} {Telescope} data'',} \textit{ Phys. Rev. Lett.}
  \textbf{ 115} (2015) 231301,
  \href{http://dx.doi.org/10.1103/PhysRevLett.115.231301}{\doi{10.1103/PhysRevLett.115.231301}},
  \href{http://www.arXiv.org/abs/1503.02641}{\texttt{arXiv:1503.02641}}.

\bibitem{Planck:2018vyg}
\hrefCMSnoop {}{{Planck} Collaboration, ``{Planck} 2018 results. {VI}.
  {C}osmological parameters'',} \textit{ Astron. Astrophys.} \textbf{ 641}
  (2020) A6,
  \href{http://dx.doi.org/10.1051/0004-6361/201833910}{\doi{10.1051/0004-6361/201833910}},
  \href{http://www.arXiv.org/abs/1807.06209}{\texttt{arXiv:1807.06209}}.
  [Erratum: \DOI{10.1051/0004-6361/201833910e}].

\bibitem{Schumann_2019}
\hrefCMSnoop {}{M.~Schumann, ``Direct detection of {WIMP} dark matter: Concepts
  and status'',} \textit{ J. Phys. G} \textbf{ 46} (2019) 103003,
  \href{http://dx.doi.org/10.1088/1361-6471/ab2ea5}{\doi{10.1088/1361-6471/ab2ea5}},
  \href{http://www.arXiv.org/abs/1903.03026}{\texttt{arXiv:1903.03026}}.

\bibitem{PhysRevLett.118.021303}
\hrefCMSnoop {}{{LUX} Collaboration, ``Results from a search for dark matter in
  the complete {LUX} exposure'',} \textit{ Phys. Rev. Lett.} \textbf{ 118}
  (2017) 021303,
  \href{http://dx.doi.org/10.1103/PhysRevLett.118.021303}{\doi{10.1103/PhysRevLett.118.021303}},
  \href{http://www.arXiv.org/abs/1608.07648}{\texttt{arXiv:1608.07648}}.

\bibitem{PhysRevD.103.063028}
\hrefCMSnoop {}{{XENON} Collaboration, ``Search for inelastic scattering of
  {WIMP} dark matter in {XENON1T}'',} \textit{ Phys. Rev. D} \textbf{ 103}
  (2021) 063028,
  \href{http://dx.doi.org/10.1103/PhysRevD.103.063028}{\doi{10.1103/PhysRevD.103.063028}},
  \href{http://www.arXiv.org/abs/2011.10431}{\texttt{arXiv:2011.10431}}.

\bibitem{SuperCDMS:2017mbc}
\hrefCMSnoop {}{{SuperCDMS} Collaboration, ``{Results from the Super Cryogenic
  Dark Matter Search Experiment at Soudan}'',} \textit{ Phys. Rev. Lett.}
  \textbf{ 120} (2018) 061802,
  \href{http://dx.doi.org/10.1103/PhysRevLett.120.061802}{\doi{10.1103/PhysRevLett.120.061802}},
  \href{http://www.arXiv.org/abs/1708.08869}{\texttt{arXiv:1708.08869}}.

\bibitem{ATLASConstraintsMediatorbasedDark2019}
\hrefCMSnoop {}{{ATLAS Collaboration}, ``Constraints on mediator-based dark
  matter and scalar dark energy models using $\sqrt{s} =$ 13 {{TeV}} pp
  collision data collected by the {{ATLAS}} detector'',} \textit{ JHEP}
  \textbf{ 05} (2019) 142,
  \href{http://dx.doi.org/10.1007/JHEP05(2019)142}{\doi{10.1007/JHEP05(2019)142}},
  \href{http://www.arXiv.org/abs/1903.01400}{\texttt{arXiv:1903.01400}}.

\bibitem{ATLAS_Monojet2021}
\hrefCMSnoop {}{{ATLAS Collaboration}, ``Search for new phenomena in events
  with an energetic jet and missing transverse momentum in $pp$ collisions at
  $\sqrt{s}=13\text{ }\text{ }\mathrm{TeV}$ with the {ATLAS} detector'',}
  \textit{ Phys. Rev. D} \textbf{ 103} (2021) 112006,
  \href{http://dx.doi.org/10.1103/PhysRevD.103.112006}{\doi{10.1103/PhysRevD.103.112006}},
  \href{http://www.arXiv.org/abs/2102.10874}{\texttt{arXiv:2102.10874}}.

\bibitem{CMS_Monojet2021}
\hrefCMSnoop {}{{CMS Collaboration}, ``Search for new particles in events with
  energetic jets and large missing transverse momentum in proton-proton
  collisions at $\sqrt{s} =$ 13 {{TeV}}'',} \textit{ JHEP} \textbf{ 11} (2021)
  153,
  \href{http://dx.doi.org/10.1007/JHEP11(2021)153}{\doi{10.1007/JHEP11(2021)153}},
  \href{http://www.arXiv.org/abs/2107.13021}{\texttt{arXiv:2107.13021}}.

\bibitem{ATLAS_MonoHiggs2021}
\hrefCMSnoop {}{{ATLAS Collaboration}, ``Search for dark matter produced in
  association with a standard model {H}iggs boson decaying into b-quarks using
  the full {R}un 2 dataset from the {ATLAS} detector'',} \textit{ JHEP}
  \textbf{ 11} (2021) 209,
  \href{http://dx.doi.org/10.1007/JHEP11(2021)209}{\doi{10.1007/JHEP11(2021)209}},
  \href{http://www.arXiv.org/abs/2108.13391}{\texttt{arXiv:2108.13391}}.

\bibitem{Atlas_HiggsInv2023}
\hrefCMSnoop {}{{ATLAS Collaboration}, ``Combination of searches for invisible
  decays of the {Higgs} boson using 139 fb$^{-1}$ of proton-proton collision
  data at $\sqrt{s}$=13 {TeV} collected with the {ATLAS} experiment'',}
  \textit{ Phys. Lett. B} \textbf{ 842} (2023) 137963,
  \href{http://dx.doi.org/10.1016/j.physletb.2023.137963}{\doi{10.1016/j.physletb.2023.137963}},
  \href{http://www.arXiv.org/abs/2301.10731}{\texttt{arXiv:2301.10731}}.

\bibitem{CMS_HiggsInv2023}
\hrefCMSnoop {}{{CMS Collaboration}, ``{A search for decays of the Higgs boson
  to invisible particles in events with a top-antitop quark pair or a vector
  boson in proton-proton collisions at $\sqrt{s}$ = 13 TeV}'',} \textit{ Eur.
  Phys. J. C} \textbf{ 83} (2023) 933,
  \href{http://dx.doi.org/10.1140/epjc/s10052-023-11952-7}{\doi{10.1140/epjc/s10052-023-11952-7}},
  \href{http://www.arXiv.org/abs/2303.01214}{\texttt{arXiv:2303.01214}}.

\bibitem{ATLAS:2018rjc}
\hrefCMSnoop {}{{ATLAS Collaboration}, ``Search for long-lived particles in
  final states with displaced dimuon vertices in $pp$ collisions at $\sqrt{s}=$
  13 {TeV} with the {ATLAS} detector'',} \textit{ Phys. Rev. D} \textbf{ 99}
  (2019) 012001,
  \href{http://dx.doi.org/10.1103/PhysRevD.99.012001}{\doi{10.1103/PhysRevD.99.012001}},
  \href{http://www.arXiv.org/abs/1808.03057}{\texttt{arXiv:1808.03057}}.

\bibitem{ATLAS:2019fwx}
\hrefCMSnoop {}{{ATLAS Collaboration}, ``Search for displaced vertices of
  oppositely charged leptons from decays of long-lived particles in $pp$
  collisions at $\sqrt {s} = 13$ {TeV} with the {ATLAS} detector'',} \textit{
  Phys. Lett. B} \textbf{ 801} (2020) 135114,
  \href{http://dx.doi.org/10.1016/j.physletb.2019.135114}{\doi{10.1016/j.physletb.2019.135114}},
  \href{http://www.arXiv.org/abs/1907.10037}{\texttt{arXiv:1907.10037}}.

\bibitem{Alimena_2020}
\hrefCMSnoop {}{J.~Alimena {et~al.}, ``Searching for long-lived particles
  beyond the standard model at the {Large Hadron Collider}'',} \textit{ J.
  Phys. G} \textbf{ 47} (2020) 090501,
  \href{http://dx.doi.org/10.1088/1361-6471/ab4574}{\doi{10.1088/1361-6471/ab4574}},
  \href{http://www.arXiv.org/abs/1903.04497}{\texttt{arXiv:1903.04497}}.

\bibitem{ATLAS_LLP_leptons}
\hrefCMSnoop {}{{ATLAS Collaboration}, ``Search for light long-lived neutral
  particles produced in pp collisions at $\sqrt{s} =$ 13 {TeV} and decaying
  into collimated leptons or light hadrons with the {ATLAS} detector'',}
  \textit{ Eur. Phys. J. C} \textbf{ 80} (2020) 450,
  \href{http://dx.doi.org/10.1140/epjc/s10052-020-7997-4}{\doi{10.1140/epjc/s10052-020-7997-4}},
  \href{http://www.arXiv.org/abs/1909.01246}{\texttt{arXiv:1909.01246}}.

\bibitem{sirunyanSearchDecaysStopped2018}
\hrefCMSnoop {}{{CMS Collaboration}, ``Search for decays of stopped exotic
  long-lived particles produced in proton-proton collisions at $\sqrt{s}=13$
  {TeV}'',} \textit{ JHEP} \textbf{ 05} (2018) 127,
  \href{http://dx.doi.org/10.1007/JHEP05(2018)127}{\doi{10.1007/JHEP05(2018)127}},
  \href{http://www.arXiv.org/abs/1801.00359}{\texttt{arXiv:1801.00359}}.

\bibitem{LHCb:2020ysn}
\hrefCMSnoop {}{{LHCb Collaboration}, ``{Searches for low-mass dimuon
  resonances}'',} \textit{ JHEP} \textbf{ 10} (2020) 156,
  \href{http://dx.doi.org/10.1007/JHEP10(2020)156}{\doi{10.1007/JHEP10(2020)156}},
  \href{http://www.arXiv.org/abs/2007.03923}{\texttt{arXiv:2007.03923}}.

\bibitem{PhysRevLett.124.131802}
\hrefCMSnoop {}{{CMS Collaboration}, ``Search for a narrow resonance lighter
  than 200 {GeV} decaying to a pair of muons in proton-proton collisions at
  $\sqrt{s}=13\text{ }\text{ }\mathrm{TeV}$'',} \textit{ Phys. Rev. Lett.}
  \textbf{ 124} (2020) 131802,
  \href{http://dx.doi.org/10.1103/PhysRevLett.124.131802}{\doi{10.1103/PhysRevLett.124.131802}},
  \href{http://www.arXiv.org/abs/1912.04776}{\texttt{arXiv:1912.04776}}.

\bibitem{CMS_displacedJets}
\hrefCMSnoop {}{{CMS Collaboration}, ``{Search for long-lived particles using
  displaced jets in proton-proton collisions at $\sqrt{s} = $ 13 TeV}'',}
  \textit{ Phys. Rev. D} \textbf{ 104} (2021) 012015,
  \href{http://dx.doi.org/10.1103/PhysRevD.104.012015}{\doi{10.1103/PhysRevD.104.012015}},
  \href{http://www.arXiv.org/abs/2012.01581}{\texttt{arXiv:2012.01581}}.

\bibitem{scoutingdispdimuon2022}
\hrefCMSnoop {}{{CMS Collaboration}, ``{Search for long-lived particles
  decaying into muon pairs in proton-proton collisions at $\sqrt{s} =$ 13 TeV
  collected with a dedicated high-rate data stream}'',} \textit{ JHEP} \textbf{
  04} (2022) 062,
  \href{http://dx.doi.org/10.1007/JHEP04(2022)062}{\doi{10.1007/JHEP04(2022)062}},
  \href{http://www.arXiv.org/abs/2112.13769}{\texttt{arXiv:2112.13769}}.

\bibitem{EXO-18-003}
\hrefCMSnoop {}{{CMS Collaboration}, ``{Search for long-lived particles
  decaying to leptons with large impact parameter in proton-proton collisions
  at $\sqrt{s}$ = 13 TeV}'',} \textit{ Eur. Phys. J. C} \textbf{ 82} (2022)
  153,
  \href{http://dx.doi.org/10.1140/epjc/s10052-022-10027-3}{\doi{10.1140/epjc/s10052-022-10027-3}},
  \href{http://www.arXiv.org/abs/2110.04809}{\texttt{arXiv:2110.04809}}.

\bibitem{EXO-21-006}
\hrefCMSnoop {}{{CMS Collaboration}, ``{Search for long-lived particles
  decaying to a pair of muons in proton-proton collisions at $ \sqrt{s} $ = 13
  TeV}'',} \textit{ JHEP} \textbf{ 05} (2023) 228,
  \href{http://dx.doi.org/10.1007/JHEP05(2023)228}{\doi{10.1007/JHEP05(2023)228}},
  \href{http://www.arXiv.org/abs/2205.08582}{\texttt{arXiv:2205.08582}}.

\bibitem{PhysRevD.43.3191}
\hrefCMSnoop {}{K.~Griest and D.~Seckel, ``Three exceptions in the calculation
  of relic abundances'',} \textit{ Phys. Rev. D} \textbf{ 43} (1991) 3191,
  \href{http://dx.doi.org/10.1103/PhysRevD.43.3191}{\doi{10.1103/PhysRevD.43.3191}}.

\bibitem{Edsj__1997}
\hrefCMSnoop {}{J.~Edsj{\"o} and P.~Gondolo, ``Neutralino relic density
  including coannihilations'',} \textit{ Phys. Rev. D} \textbf{ 56} (1997)
  1879,
  \href{http://dx.doi.org/10.1103/physrevd.56.1879}{\doi{10.1103/physrevd.56.1879}},
  \href{http://www.arXiv.org/abs/hep-ph/9704361}{\texttt{arXiv:hep-ph/9704361}}.

\bibitem{izaguirre}
\hrefCMSnoop {}{E.~Izaguirre, G.~Krnjaic, and B.~Shuve, ``Discovering inelastic
  thermal-relic dark matter at colliders'',} \textit{ Phys. Rev. D} \textbf{
  93} (2016) 063523,
  \href{http://dx.doi.org/10.1103/PhysRevD.93.063523}{\doi{10.1103/PhysRevD.93.063523}},
  \href{http://www.arXiv.org/abs/1508.03050}{\texttt{arXiv:1508.03050}}.

\bibitem{berlinInelasticDarkMatter2019}
\hrefCMSnoop {}{A.~Berlin and F.~Kling, ``Inelastic dark matter at the {LHC}
  lifetime frontier: {ATLAS}, {CMS}, {LHCb}, {CODEX}-b, {FASER}, and
  {MATHUSLA}'',} \textit{ Phys. Rev. D} \textbf{ 99} (2019) 015021,
  \href{http://dx.doi.org/10.1103/PhysRevD.99.015021}{\doi{10.1103/PhysRevD.99.015021}},
  \href{http://www.arXiv.org/abs/1810.01879}{\texttt{arXiv:1810.01879}}.

\bibitem{Tucker-Smith:2001myb}
\hrefCMSnoop {}{D.~Smith and N.~Weiner, ``Inelastic dark matter'',} \textit{
  Phys. Rev. D} \textbf{ 64} (2001) 043502,
  \href{http://dx.doi.org/10.1103/PhysRevD.64.043502}{\doi{10.1103/PhysRevD.64.043502}},
  \href{http://www.arXiv.org/abs/hep-ph/0101138}{\texttt{arXiv:hep-ph/0101138}}.

\bibitem{Holdom:1985ag}
\hrefCMSnoop {}{B.~Holdom, ``Two {U(1)'s} and epsilon charge shifts'',}
  \textit{ Phys. Lett. B} \textbf{ 166} (1986) 196,
  \href{http://dx.doi.org/10.1016/0370-2693(86)91377-8}{\doi{10.1016/0370-2693(86)91377-8}}.

\bibitem{Batell_2014}
\hrefCMSnoop {}{B.~Batell, R.~Essig, and Z.~Surujon, ``Strong constraints on
  sub-{GeV} dark sectors from {SLAC} beam dump {E137}'',} \textit{ Phys. Rev.
  Lett.} \textbf{ 113} (2014) 171802,
  \href{http://dx.doi.org/10.1103/physrevlett.113.171802}{\doi{10.1103/physrevlett.113.171802}},
  \href{http://www.arXiv.org/abs/1406.2698}{\texttt{arXiv:1406.2698}}.

\bibitem{PhysRevD.96.055007}
\hrefCMSnoop {}{E.~Izaguirre, Y.~Kahn, G.~Krnjaic, and M.~Moschella, ``Testing
  light dark matter coannihilation with fixed-target experiments'',} \textit{
  Phys. Rev. D} \textbf{ 96} (2017) 055007,
  \href{http://dx.doi.org/10.1103/PhysRevD.96.055007}{\doi{10.1103/PhysRevD.96.055007}},
  \href{http://www.arXiv.org/abs/1703.06881}{\texttt{arXiv:1703.06881}}.

\bibitem{Mongillo_2023}
M.~Mongillo\hrefCMSnoop {}{ {et~al.}, ``Constraining light thermal inelastic
  dark matter with {NA}64'',} \textit{ Eur. Phys. J. C} \textbf{ 83} (2023)
  391,
  \href{http://dx.doi.org/10.1140/epjc/s10052-023-11536-5}{\doi{10.1140/epjc/s10052-023-11536-5}},
  \href{http://www.arXiv.org/abs/2302.05414}{\texttt{arXiv:2302.05414}}.

\bibitem{PhysRevD.38.3375}
J.~D. Bjorken\hrefCMSnoop {}{ {et~al.}, ``Search for neutral metastable
  penetrating particles produced in the {SLAC} beam dump'',} \textit{ Phys.
  Rev. D} \textbf{ 38} (1988) 3375,
  \href{http://dx.doi.org/10.1103/PhysRevD.38.3375}{\doi{10.1103/PhysRevD.38.3375}}.

\bibitem{Auerbach_2001}
\hrefCMSnoop {}{{LSND} Collaboration, ``Measurement of electron-neutrino
  electron elastic scattering'',} \textit{ Phys. Rev. D} \textbf{ 63} (2001)
  112001,
  \href{http://dx.doi.org/10.1103/PhysRevD.63.112001}{\doi{10.1103/PhysRevD.63.112001}},
  \href{http://www.arXiv.org/abs/hep-ex/0101039}{\texttt{arXiv:hep-ex/0101039}}.

\bibitem{Hook:2010tw}
\hrefCMSnoop {}{A.~Hook, E.~Izaguirre, and J.~G. Wacker, ``Model independent
  bounds on kinetic mixing'',} \textit{ Adv. High Energy Phys.} \textbf{ 2011}
  (2011) 859762,
  \href{http://dx.doi.org/10.1155/2011/859762}{\doi{10.1155/2011/859762}},
  \href{http://www.arXiv.org/abs/1006.0973}{\texttt{arXiv:1006.0973}}.

\bibitem{Lees_2017}
\hrefCMSnoop {}{{BaBar} Collaboration, ``Search for invisible decays of a dark
  photon produced in ${e}^{+}{e}^{\ensuremath{-}}$ collisions at {BaBar}'',}
  \textit{ Phys. Rev. Lett.} \textbf{ 119} (2017) 131804,
  \href{http://dx.doi.org/10.1103/physrevlett.119.131804}{\doi{10.1103/physrevlett.119.131804}},
  \href{http://www.arXiv.org/abs/1702.03327}{\texttt{arXiv:1702.03327}}.

\bibitem{Baer_2011}
\hrefCMSnoop {}{H.~Baer, V.~Barger, and P.~Huang, ``Hidden {SUSY} at the {LHC}:
  the light {higgsino}-world scenario and the role of a lepton collider'',}
  \textit{ JHEP} \textbf{ 11} (2011) 031,
  \href{http://dx.doi.org/10.1007/jhep11(2011)031}{\doi{10.1007/jhep11(2011)031}},
  \href{http://www.arXiv.org/abs/1107.5581}{\texttt{arXiv:1107.5581}}.

\bibitem{PhysRevD.89.075007}
\hrefCMSnoop {}{Z.~Han, G.~D. Kribs, A.~Martin, and A.~Menon, ``Hunting
  quasidegenerate {Higgsinos}'',} \textit{ Phys. Rev. D} \textbf{ 89} (2014)
  075007,
  \href{http://dx.doi.org/10.1103/PhysRevD.89.075007}{\doi{10.1103/PhysRevD.89.075007}},
  \href{http://www.arXiv.org/abs/1401.1235}{\texttt{arXiv:1401.1235}}.

\bibitem{hepdata}
\hrefCMSnoop {}{}{HEPD}ata record for this analysis, 2023.
\newblock
  \href{http://dx.doi.org/10.17182/hepdata.140434}{\doi{10.17182/hepdata.140434}}.

\bibitem{atlascollaboration2022search}
\hrefCMSnoop {}{{ATLAS Collaboration}, ``Search for heavy neutral leptons in
  decays of {$W$} bosons using a dilepton displaced vertex in $\sqrt{s}=13$
  {TeV} $pp$ collisions with the {ATLAS} detector'',} \textit{ Phys. Rev.
  Lett.} \textbf{ 131} (2023) 061803,
  \href{http://dx.doi.org/10.1103/PhysRevLett.131.061803}{\doi{10.1103/PhysRevLett.131.061803}},
  \href{http://www.arXiv.org/abs/2204.11988}{\texttt{arXiv:2204.11988}}.

\bibitem{Chatrchyan:2008zzk}
\hrefCMSnoop {}{{CMS Collaboration}, ``The {CMS} experiment at the {CERN
  LHC}'',} \textit{ JINST} \textbf{ 3} (2008) S08004,
  \href{http://dx.doi.org/10.1088/1748-0221/3/08/S08004}{\doi{10.1088/1748-0221/3/08/S08004}}.

\bibitem{CMS-PRF-14-001}
\hrefCMSnoop {}{{CMS Collaboration}, ``Particle-flow reconstruction and global
  event description with the {CMS} detector'',} \textit{ JINST} \textbf{ 12}
  (2017) P10003,
  \href{http://dx.doi.org/10.1088/1748-0221/12/10/P10003}{\doi{10.1088/1748-0221/12/10/P10003}},
  \href{http://www.arXiv.org/abs/1706.04965}{\texttt{arXiv:1706.04965}}.

\bibitem{MadGraph}
J.~Alwall\hrefCMSnoop {}{ {et~al.}, ``The automated computation of tree-level
  and next-to-leading order differential cross sections, and their matching to
  parton shower simulations'',} \textit{ JHEP} \textbf{ 07} (2014) 079,
  \href{http://dx.doi.org/10.1007/JHEP07(2014)079}{\doi{10.1007/JHEP07(2014)079}},
  \href{http://www.arXiv.org/abs/1405.0301}{\texttt{arXiv:1405.0301}}.

\bibitem{Alwall_2007}
J.~Alwall\hrefCMSnoop {}{ {et~al.}, ``Comparative study of various algorithms
  for the merging of parton showers and matrix elements in hadronic
  collisions'',} \textit{ Eur. Phys. J. C} \textbf{ 53} (2007) 473,
  \href{http://dx.doi.org/10.1140/epjc/s10052-007-0490-5}{\doi{10.1140/epjc/s10052-007-0490-5}},
  \href{http://www.arXiv.org/abs/0706.2569}{\texttt{arXiv:0706.2569}}.

\bibitem{Sjostrand:2014zea}
T.~Sj{\"o}strand\hrefCMSnoop {}{ {et~al.}, ``An introduction to {PYTHIA}
  8.2'',} \textit{ Comput. Phys. Commun.} \textbf{ 191} (2015) 159,
  \href{http://dx.doi.org/10.1016/j.cpc.2015.01.024}{\doi{10.1016/j.cpc.2015.01.024}},
  \href{http://www.arXiv.org/abs/1410.3012}{\texttt{arXiv:1410.3012}}.

\bibitem{khachatryanEventGeneratorTunes2016}
\hrefCMSnoop {}{{CMS Collaboration}, ``Event generator tunes obtained from
  underlying event and multiparton scattering measurements'',} \textit{ Eur.
  Phys. J. C} \textbf{ 76} (2016) 155,
  \href{http://dx.doi.org/10.1140/epjc/s10052-016-3988-x}{\doi{10.1140/epjc/s10052-016-3988-x}},
  \href{http://www.arXiv.org/abs/1512.00815}{\texttt{arXiv:1512.00815}}.

\bibitem{sirunyanExtractionValidationNew2020}
\hrefCMSnoop {}{{CMS Collaboration}, ``Extraction and validation of a new set
  of {CMS PYTHIA8} tunes from underlying-event measurements'',} \textit{ Eur.
  Phys. J. C} \textbf{ 80} (2020) 4,
  \href{http://dx.doi.org/10.1140/epjc/s10052-019-7499-4}{\doi{10.1140/epjc/s10052-019-7499-4}},
  \href{http://www.arXiv.org/abs/1903.12179}{\texttt{arXiv:1903.12179}}.

\bibitem{ballPartonDistributionsLHC2015}
R.~D. Ball\hrefCMSnoop {}{ {et~al.}, ``Parton distributions for the {LHC Run
  II}'',} \textit{ JHEP} \textbf{ 04} (2015) 40,
  \href{http://dx.doi.org/10.1007/JHEP04(2015)040}{\doi{10.1007/JHEP04(2015)040}},
  \href{http://www.arXiv.org/abs/1410.8849}{\texttt{arXiv:1410.8849}}.

\bibitem{ballPartonDistributionsHighprecision2017}
R.~D. Ball\hrefCMSnoop {}{ {et~al.}, ``Parton distributions from high-precision
  collider data'',} \textit{ Eur. Phys. J. C} \textbf{ 77} (2017) 663,
  \href{http://dx.doi.org/10.1140/epjc/s10052-017-5199-5}{\doi{10.1140/epjc/s10052-017-5199-5}},
  \href{http://www.arXiv.org/abs/1706.00428}{\texttt{arXiv:1706.00428}}.

\bibitem{Agostinelli:2002hh}
\hrefCMSnoop {}{{GEANT4} Collaboration, ``{\GEANTfour}\textemdash a simulation
  toolkit'',} \textit{ Nucl. Instrum. Methods Phys. Res., Sect. A} \textbf{
  506} (2003) 250,
  \href{http://dx.doi.org/10.1016/S0168-9002(03)01368-8}{\doi{10.1016/S0168-9002(03)01368-8}}.

\bibitem{Khachatryan:2016bia}
\hrefCMSnoop {}{{CMS Collaboration}, ``The {CMS} trigger system'',} \textit{
  JINST} \textbf{ 12} (2017) P01020,
  \href{http://dx.doi.org/10.1088/1748-0221/12/01/P01020}{\doi{10.1088/1748-0221/12/01/P01020}},
\href{http://www.arXiv.org/abs/1609.02366}{\texttt{arXiv:1609.02366}}.

\bibitem{Sirunyan:2298594}
\hrefCMSnoop {}{{CMS Collaboration}, ``Identification of heavy-flavour jets
  with the {CMS} detector in pp collisions at 13 {TeV}'',} \textit{ JINST}
  \textbf{ 13} (2017) P05011,
  \href{http://dx.doi.org/10.1088/1748-0221/13/05/P05011}{\doi{10.1088/1748-0221/13/05/P05011}},
  \href{http://www.arXiv.org/abs/1712.07158}{\texttt{arXiv:1712.07158}}.

\bibitem{CMS-DP-2023-005}
\href {https://cds.cern.ch/record/2854609}{{CMS Collaboration}, ``Performance
  summary of {AK4} jet b tagging with data from proton-proton collisions at 13
  {TeV} with the {CMS} detector'',} CMS Detector Performance Note
  CMS-DP-2023-005, 2023.

\bibitem{Sirunyan_2018}
\hrefCMSnoop {}{{CMS Collaboration}, ``Performance of the {CMS} muon detector
  and muon reconstruction with proton-proton collisions at $\sqrt{s} = $ 13
  {TeV}'',} \textit{ JINST} \textbf{ 13} (2018) P06015,
  \href{http://dx.doi.org/10.1088/1748-0221/13/06/P06015}{\doi{10.1088/1748-0221/13/06/P06015}},
  \href{http://www.arXiv.org/abs/1804.04528}{\texttt{arXiv:1804.04528}}.

\bibitem{fruhwirthApplicationKalmanFiltering1987a}
\hrefCMSnoop {}{R.~Fr{\"u}hwirth, ``Application of {Kalman} filtering to track
  and vertex fitting'',} \textit{ Nucl. Instrum. Methods Phys. Res., Sect. A}
  \textbf{ 262} (1987) 444,
  \href{http://dx.doi.org/10.1016/0168-9002(87)90887-4}{\doi{10.1016/0168-9002(87)90887-4}}.

\bibitem{cmscollaborationSearchLonglivedParticles2019}
\hrefCMSnoop {}{{CMS Collaboration}, ``Search for long-lived particles using
  delayed photons in proton-proton collisions at $\sqrt{s} =$ 13 {TeV}'',}
  \textit{ Phys. Rev. D} \textbf{ 100} (2019) 112003,
  \href{http://dx.doi.org/10.1103/PhysRevD.100.112003}{\doi{10.1103/PhysRevD.100.112003}},
  \href{http://www.arXiv.org/abs/1909.06166}{\texttt{arXiv:1909.06166}}.

\bibitem{Sirunyan:2759951}
\hrefCMSnoop {}{{CMS Collaboration}, ``Precision luminosity measurement in
  proton-proton collisions at $\sqrt{s} = 13$ {TeV} in 2015 and 2016 at
  {CMS}'',} \textit{ Eur. Phys. J. C} \textbf{ 81} (2021) 800,
  \href{http://dx.doi.org/10.1140/epjc/s10052-021-09538-2}{\doi{10.1140/epjc/s10052-021-09538-2}},
  \href{http://www.arXiv.org/abs/2104.01927}{\texttt{arXiv:2104.01927}}.

\bibitem{CMS-PAS-LUM-17-004}
\href {https://cds.cern.ch/record/2621960}{{CMS Collaboration}, ``{CMS}
  luminosity measurement for the 2017 data-taking period at $\sqrt{s} =
  13~\mathrm{TeV}$'',} CMS Physics Analysis Summary CMS-PAS-LUM-17-004, 2018.

\bibitem{CMS-PAS-LUM-18-002}
\href {https://cds.cern.ch/record/2676164}{{CMS Collaboration}, ``{CMS}
  luminosity measurement for the 2018 data-taking period at $\sqrt{s} = 13$
  {TeV}'',} CMS Physics Analysis Summary CMS-PAS-LUM-18-002, 2019.

\bibitem{JUNK1999435}
\hrefCMSnoop {}{T.~Junk, ``Confidence level computation for combining searches
  with small statistics'',} \textit{ Nucl. Instrum. Methods Phys. Res., Sect.
  A} \textbf{ 434} (1999) 435,
  \href{http://dx.doi.org/10.1016/S0168-9002(99)00498-2}{\doi{10.1016/S0168-9002(99)00498-2}},
  \href{http://www.arXiv.org/abs/hep-ex/9902006}{\texttt{arXiv:hep-ex/9902006}}.

\bibitem{Read_2002}
\hrefCMSnoop {}{A.~L. Read, ``Presentation of search results: The
  {CL}$_{\text{s}}$ technique'',} \textit{ J. Phys. G} \textbf{ 28} (2002)
  2693,
  \href{http://dx.doi.org/10.1088/0954-3899/28/10/313}{\doi{10.1088/0954-3899/28/10/313}}.

\end{thebibliography}\endgroup
\cleardoublepage \appendix\section{The CMS Collaboration \label{app:collab}}\begin{sloppypar}\hyphenpenalty=5000\widowpenalty=500\clubpenalty=5000
\cmsinstitute{Yerevan Physics Institute, Yerevan, Armenia}
{\tolerance=6000
A.~Hayrapetyan, A.~Tumasyan\cmsAuthorMark{1}\cmsorcid{0009-0000-0684-6742}
\par}
\cmsinstitute{Institut f\"{u}r Hochenergiephysik, Vienna, Austria}
{\tolerance=6000
W.~Adam\cmsorcid{0000-0001-9099-4341}, J.W.~Andrejkovic, T.~Bergauer\cmsorcid{0000-0002-5786-0293}, S.~Chatterjee\cmsorcid{0000-0003-2660-0349}, K.~Damanakis\cmsorcid{0000-0001-5389-2872}, M.~Dragicevic\cmsorcid{0000-0003-1967-6783}, A.~Escalante~Del~Valle\cmsorcid{0000-0002-9702-6359}, P.S.~Hussain\cmsorcid{0000-0002-4825-5278}, M.~Jeitler\cmsAuthorMark{2}\cmsorcid{0000-0002-5141-9560}, N.~Krammer\cmsorcid{0000-0002-0548-0985}, D.~Liko\cmsorcid{0000-0002-3380-473X}, I.~Mikulec\cmsorcid{0000-0003-0385-2746}, J.~Schieck\cmsAuthorMark{2}\cmsorcid{0000-0002-1058-8093}, R.~Sch\"{o}fbeck\cmsorcid{0000-0002-2332-8784}, D.~Schwarz\cmsorcid{0000-0002-3821-7331}, M.~Sonawane\cmsorcid{0000-0003-0510-7010}, S.~Templ\cmsorcid{0000-0003-3137-5692}, W.~Waltenberger\cmsorcid{0000-0002-6215-7228}, C.-E.~Wulz\cmsAuthorMark{2}\cmsorcid{0000-0001-9226-5812}
% ...
\par}
\vskip\cmsinstskip
\dag:~Deceased\\
$^{1}$Also at Yerevan State University, Yerevan, Armenia\\
$^{2}$Also at TU Wien, Vienna, Austria\\
% ...
\end{sloppypar}
\end{document}